\documentclass[10pt]{article}

\usepackage{amsmath}
\usepackage{amssymb}

\usepackage{cite}

\usepackage{hyperref}
\usepackage{breakurl}

\usepackage{lineno}

\usepackage{microtype}
\DisableLigatures[f]{encoding = *, family = * }

\usepackage{graphicx,color}

\usepackage[labelfont=bf,labelsep=period,justification=raggedright]{caption}

\makeatletter
\renewcommand{\@biblabel}[1]{\quad#1.}
\makeatother

\date{}

\newcommand{\selfarrow}{
\; \raisebox{-2pt}{\rotatebox{90}{\large $\circlearrowleft$}}
}

\begin{document}

\begin{flushleft}
{\Large
\textbf{Optimal census by quorum sensing}
}
\\
Thibaud Taillefumier$^{1}$, 
Ned S. Wingreen$^{1,2,\ast}$
\\
\textbf{1}  Lewis-Sigler Institute for Integrative Genomics, Princeton University, Princeton, NJ 08544, USA
\\
\textbf{2} Department of Molecular Biology, Princeton University, Princeton, NJ 08544, USA
\\
$\ast$ Email: \href{mailto:wingreen@princeton.edu}{wingreen@princeton.edu}
\end{flushleft}


\section*{Abstract}

Quorum sensing is the regulation of gene expression in response to changes in cell density.
To measure their cell density, bacterial populations produce and detect diffusible molecules called autoinducers.
Individual bacteria  internally represent the external concentration of autoinducers via the level of monitor proteins.
In turn, these monitor proteins typically regulate both their own production and the production of autoinducers, thereby establishing internal and external feedbacks.
Here, we ask whether feedbacks can increase the information available to cells about their local density.
We quantify available information as the mutual information between the abundance of a monitor protein and the local cell density for biologically relevant models of quorum sensing.
Using variational methods, we demonstrate that feedbacks can increase information transmission, allowing bacteria to resolve up to two additional ranges of cell density when compared with bistable quorum-sensing systems.
Our analysis is relevant to multi-agent systems that track an external driver implicitly via an endogenously generated signal.


\section*{Author Summary}

Bacteria regulate gene expression in response to changes in cell density in a process called quorum sensing. 
To synchronize their gene-expression programs, these bacteria need to glean as much information as possible about their cell density. 
Our study is the first to physically model the flow of information in a quorum-sensing microbial community, wherein the internal regulator of the individualsÕ response tracks the external cell density via an endogenously generated shared signal. 
Combining information theory and Lagrangian formalism, we find that quorum-sensing systems can improve their information capabilities by tuning circuit feedbacks. 
Our analysis suggests that achieving information benefit via feedback requires dedicated systems to control gene expression noise, such as sRNA-based regulation.


\section*{Introduction}

%
%
%
To successfully colonize an environment, many bacteria engage in collective tasks, for instance the synthesis of a biofilm matrix or the secretion of enzymes or virulence factors.
In general, such tasks can only be performed efficiently in a prescribed temporal order and at high enough cell density.
Bacteria achieve the necessary level of coordination through quorum sensing, whereby individual cells monitor local cell density to synchronize their programs of gene expression~\cite{Miller:2001oq,Fuqua:1996bh,Nadal-Jimenez:2012hc}. 
In quorum sensing, bacteria infer cell density by producing and detecting freely diffusing molecules called autoinducers (AIs).
As cells grow and divide, the increasing external AI concentration constitutes a shared signal at the population level.
To represent this signal within each cell, the quorum-sensing circuit regulates the abundance of one or more internal monitor proteins (MPs).
In turn, these MPs act as regulators of gene expression, often inducing the genes responsible for AI production, hence the term ``autoinducer''~\cite{Ng:2009nx,Pompeani:2008}. 
Autoinduction thus establishes a positive feedback loop that can lead to switching at the population level between two stable states of gene expression, e.g. as observed in the symbiotic bioluminescent marine bacterium \emph{Vibrio fisherii}~\cite{Angeli:2004,Kramer:2005}. 
However, negative feedbacks from MP expression to AI detection are present in a related bioluminescent bacterium, \emph{Vibrio harveyi}, which exhibits a \emph{graded} quorum-sensing response~\cite{Long:2009dq,Teng:2010cr,Teng:2011fk}. 
While \emph{V. fisherii} primarily alternates between planktonic and symbiotic states~\cite{Ruby:1996fk},  quorum sensing in \emph{V. harveyi}  implements multiple --- at least three ---  states of gene expression during host infections~\cite{Henke:2004uq,Anetzberger:2009vn,Austin:2006fk,Yang:2014uq}.
To optimize their program of gene expression during cycles of colonization, bacteria such as \emph{V. harveyi} need to glean as much information as possible from AI concentration.
Here we address the question: can feedbacks from MP expression to AI production and to AI detection increase the information available to cells about their local density?\\
%
%
%
%
%
%
\indent  A natural way to quantify information transfer in quorum sensing is via the concept of mutual information (MI). 
The MI between two random variables provides a general measure of their statistical dependence. 
When evaluated between an input and output variable, the MI quantifies the amount of information, in bits, that the output conveys about the input~\cite{Shannon:1948fk,Cover:1991}.
In the context of quorum sensing, the fidelity of information processing can be quantified via the MI between cell density and the abundance of an internal monitor protein.
Biologically, we interpret this information as the number of distinct cell-density ranges that a bacterium can resolve by reading out its MP's abundance, though how bacteria utilize the available information about cell density may be complex.
For example, a bistable quorum-sensing system that only discriminates between high and low cell density can transmit at most one bit of MI. 
In contrast, bacteria with a graded quorum-sensing response can resolve more than one bit, thus enabling more than two differentiated cell-density stages.\\
%
%
%
%
%
%
\indent Here, we formulate the quorum-sensing circuit as an information channel that encodes cell density in the abundance of an internal MP.
We then optimize the MI between cell density and the MP by varying the feedbacks from MP expression to AI production and to AI detection.
We consider each bacterium as an imperfect detector and quantify its private information about cell density.
For biologically relevant models of quorum sensing, optimizing feedback approximatively doubles the  information available to a cell, providing a justification for the increased complexity of the quorum-sensing circuit required to implement feedback.
Our findings about the role of feedbacks in promoting information transfer can be understood intuitively.
External feedback allows bacteria to adjust the shared AI input to match the cells' detection capabilities, preferentially exploiting AI concentration ranges where detection is most sensitive.
Internal feedback allows a bacterium to adjust its quorum-sensing response time to achieve an optimal trade off between output noise reduction and signal tracking ability.


\section*{Results}


We imagine a bacterial population colonizing a surface, such as a clonal patch of \emph{V. harveyi} forming a biofilm~\cite{Pai:2009fu,Nadell:2013}. 
We model AI diffusion in a volume with length scale $L=100\mu \mathrm{m}$, a typical biofilm dimension.
Schematic Fig.~1A shows this volume $V$ at time $t$ when it contains $N_t$ bacteria and $A_t$ freely diffusing AI molecules, thus defining the cell density $\rho_t = N_t/V$ and AI concentration $a_t=A_t/V$.
Quorum sensing implies that, in each bacterium $i$, $0 \leq i < N_t$, the MP abundance, defined as the intracellular MP concentration $m_{i,t}$, somehow tracks the evolution of the AI concentration $a_t$.
We consider that the MP abundances $m_{i,t}$, which differ among bacteria, follow the same statistics and we refer to the MP abundance in a representative cell as $m_t$.
Assuming large numbers of molecules and cells, we adopt continuous descriptions of $\rho_t$, $a_t$, and $m_t$.
In practice, the quorum-sensing system is only responsive to cell densities for which the AI concentration lies between the receptors' detection threshold $a_-$ and saturation threshold $a_+$.
We therefore assume that the bacteria begin to engage in collective behaviors at the cell density $\rho_-$ at which the AI concentration reaches $a_-$, while quorum sensing becomes insensitive above the cell density $\rho_+$ at which the AI concentration reaches $a_+$.
Over this range, the quorum-sensing circuit raises the MP concentration from a basal level $m_-$ to a peak level $m_+$.
We treat the extremal values of $\rho_t$, $a_t$, and $m_t$ as fixed boundary conditions set by a combination of physical, chemical, and biological constraints.\\

\subsection*{Quorum-sensing model}

%
%
%
%
\indent  What specifies the input distribution $p(\rho)$ for cell densities? 
Intuitively, $p(\rho)$ represents the likelihood for a bacterium to find itself at density $\rho$, while engaged in quorum sensing.
As depicted in Fig.~1B, this suggests we identify $p(\rho)$ as the fraction of time that $\rho_t$ spends at cell density $\rho$, averaged over many cycles of colonization, growth, and dispersal. 
For simplicity, we consider that the cell density evolves continuously as a deterministic exponential function $\rho_t = \lambda(t) \propto e^{\gamma t}$, over growth periods of a single duration $T$. The boundary conditions $\lambda(0) = \rho_-$ and $\lambda(T) = \rho_+$  
specifies the growth rate $\gamma = \log (\rho_+/\rho_-) / T$ so that the cell-density time course is
\begin{eqnarray}
\lambda(t) = \rho_- \left( \frac{\rho_+}{\rho_-} \right)^{t/T} \, .
\end{eqnarray}
We specify the input distribution $p(\rho)$ as the transform of the uniform distribution $dt/T$ via the deterministic function $\lambda(t)$, i.e. by writing $p(\rho) \, d\rho = p(\rho) \lambda'(t) \, dt = dt/T$.
Thus, $p(\rho)$ is determined as $p(\rho) = 1/\big(T \lambda'(\lambda^{-1}(\rho))\big)$, where $\lambda'$ is the time derivative of the exponential time course and $\lambda^{-1}$ is the inverse of the function $\lambda$, which implies $p(\rho) \propto 1/\rho$.
Similarly, the distribution of AI concentrations $q(c)$ as well as the distribution of MP abundance $q(m)$ are  defined over $(a_-,a_+)$ and $(m_-,m_+)$ as the fraction of time that $a_t$ and $m_t$ spend in the vicinity of $a$ and $m$.
Given a time course for colony growth, the output distributions $q(c)$ and $q(m)$  are determined by the coupled dynamics of AI concentration $a_t$ and MP concentrations $m_{i,t}$.
Fig.~1C  depicts the interactions between fluctuating variables  $a_t$ and $\lbrace m_{i,t} \rbrace$, driven by bacterial growth $\rho_t$,  represented schematically as $\rho \rightarrow a \leftrightarrows m \selfarrow$. 
The figure highlights that the output rate of AIs depends on the AI concentration via the MP abundance, thus establishing an external feedback loop in addition to an internal feedback loop of self-regulated MP production.\\
%
%
%
%
%
%
\indent To specify the dynamics of the quorum-sensing response, we model the evolution of $a_t$ and $\lbrace m_{i,t} \rbrace$, $0\leq i < N_t$, through stochastic differential equations~\cite{Karatzas,Gardiner}:
\begin{eqnarray}
\label{eq:qsStoch1}
\frac{a_t}{\tau_a} =  \rho_t \, \Big \langle f_{\mathrm{ext}}(m_{i,t}) \Big \rangle_i    \, ,
\end{eqnarray}
\vspace{-10pt}
\begin{eqnarray}
\label{eq:qsStoch2}
dm_{i,t} = \left(-\frac{m_{i,t}}{\tau_m} + f_m(a_t, m_{i,t}) \right)\, dt + \sqrt{2} \sigma_m \, dW^{(i)}_t \, ,
\end{eqnarray}
where the angular brackets denote an average over the population of bacteria (see S1 Text).
Equation  \eqref{eq:qsStoch1} describes the evolution of the AI concentration $a_t$ in response to the AI production of the whole colony.
The set of equations \eqref{eq:qsStoch2} characterizes the accumulation of MPs inside each bacterium in the colony.
The function $f_{\mathrm{ext}}$ is the output rate for AI in molecules per second per cell, while $f_m$ is the output rate for  MP in molecules per second per cell volume.
We model MP self-regulation via the dependence of $f_m$ on the MP abundance $m$, which characterizes the internal feedback.
To be concrete, we consider that the MP output rate $f_m$ is proportional to the probability that the MPs bind some regulatory site times a bare MP expression rate, which is independently regulated by AI detection.
Thus,  $f_m(a_t, m_{i,t})=f_m^{(1)}(a_t)f_{\mathrm{int}}(m_{i,t})$ where $f_\mathrm{int}$ quantifies the level of self-induction $(f_{\mathrm{int}}>1)$ or self-repression $(f_{\mathrm{int}}<1)$ and $f_m^{(1)}$ is the bare output rate in the absence of MP-mediated feedback $(f_{\mathrm{int}}=1)$.
The constant $\tau_a$ denotes the time for AIs to diffuse out of the volume $V$, while $\tau_m$ denotes the lifetime of the MP set by active degradation or dilution by cell growth. 
Very generally, the timescale $\tau_m \approx 30\mathrm{min}$ will be long compared with the correlation time of the AI concentration $\tau_a  \approx L^{2}/D \approx 10\mathrm{s}$ (taking $D \approx 10^3 \mu \mathrm{m}^2/\mathrm{s}$ as the diffusion constant of autoinducers).
This separation of timescales $\tau_a \ll \tau_m$ effectively eliminates the AI-output noise as a source of stochasticity because the fast fluctuations of the AI concentration average out over the time $\tau_m$ (see S1 Text).
The MP-output noise is modeled via independent Gaussian white noise $dW^{(i)}$, with noise coefficient $\sigma_m$ considered as a function of $a_t$ and $m_t$. 
Thus, the MP abundance $m$ results from a nonlinear temporal averaging of the AI concentration $a$ over the MP lifetime, distorted by the internal noise in gene expression.
Together, the functions $f_{\mathrm{ext}}$, $f_m= f_m^{(1)} f_{\mathrm{int}}$, and the noise function $\sigma_m$ specify our quorum-sensing model.
We only consider $f_{\mathrm{ext}}$ and $f_\mathrm{int}$ as variables of optimization, hereafter referred to as the external feedback and internal feedback  (green arrows in Fig.~1C and Fig.~1D), respectively.\\
\indent For clarity, we recap the biologically relevant assumptions upon which our dynamical model relies:
$(i)$ the collective production of fast diffusing AI molecules gives rise to a homogeneous external AI concentration, $(ii)$ the fast fluctuations of the internal AI signal time average over the MP lifetime, and finally, $(iii)$ the feedback mechanisms are also fast with respect to the MP fluctuations.
Taken together, assumptions $(i)$ and $(ii)$ justify considering the internal MP expression noise as the dominant source of stochasticity.
Assumptions $(ii)$ and $(iii)$ justify the simple dependence of the expression noise function $\sigma_m$ and the feedback functions $f_\mathrm{ext}$ and $f_\mathrm{int}$ on the level of MP expression.

\begin{figure}[h]
\includegraphics[width=\textwidth]{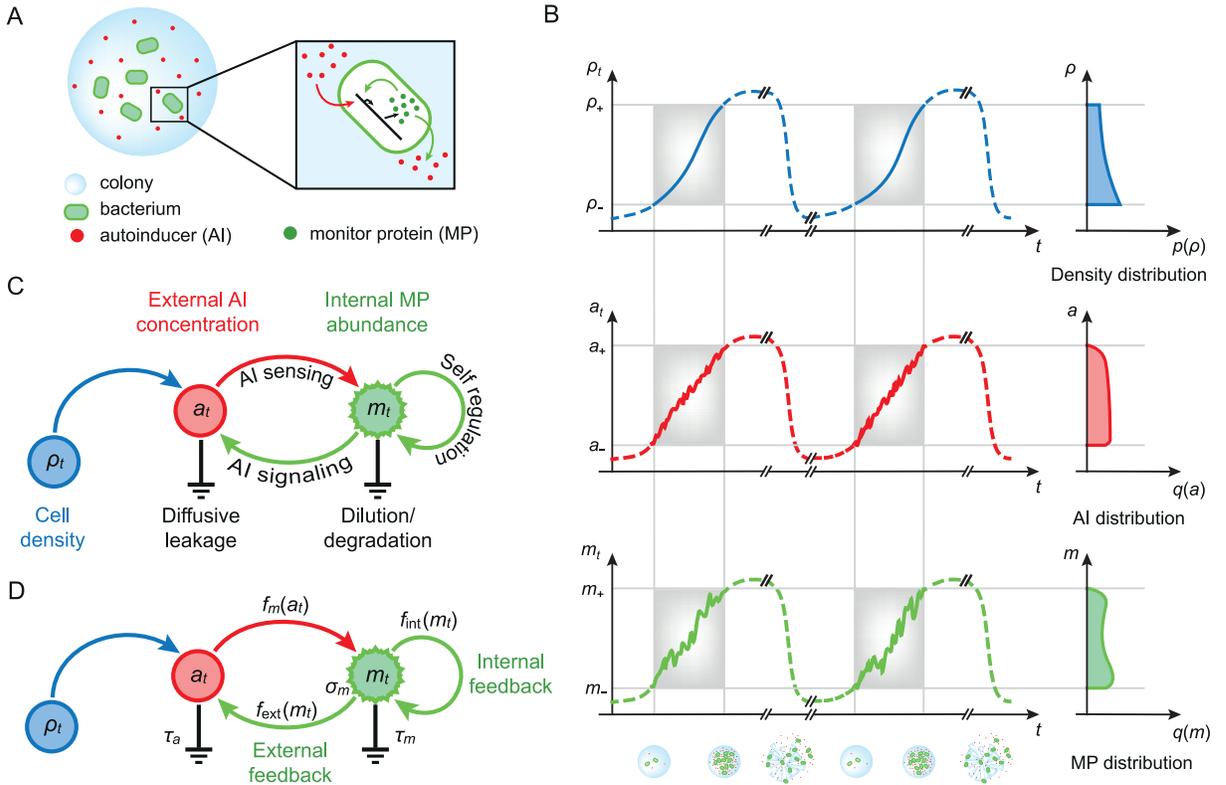}
\caption{
\textbf{Signals, statistics, and dynamics  in quorum sensing.}
\textbf{A.} A growing bacterial colony tracks the concentration of endogenously produced autoinducers (AIs) via the internal abundance of a monitor protein (MP). \textbf{B.} The quorum-sensing system responds to changing cell density $\rho$ by regulating the concentration of external autoinducers $a_t$ and the internal concentration of  MPs $m_t$.  A succession of identical growth cycles yields stationary distributions for $\rho$, $a$, and $m$ within the operational range of quorum sensing (shaded frames). \textbf{C.} Dependency graph modeling quorum sensing with feedback from MP expression to AI signaling and to AI sensing. 
\textbf{D.} Functions parametrizing the stochastic dynamics of the quorum-sensing response. The AI output rate $f_{\mathrm{ext}}$ and the self-regulation level $f_\mathrm{int}$ are emphasized as the external and internal feedback, respectively. 
}
\end{figure}

\subsection*{Quorum-sensing information channel}

%
%
%
 \indent The coupled dynamics of $\rho_t$, $a_t$, and $\lbrace m_{i,t} \rbrace$ allow us to specify the output distributions $q(c)$ and $q(m)$ as shown schematically in Fig.~1B.
These dynamics also prescribe the processing of information by each bacterium's quorum-sensing circuit.
Thus our model defines an information channel that transforms input cell density into output MP abundance according to a specific encoding scheme.
To see this, we consider equations \eqref{eq:qsStoch1} and \eqref{eq:qsStoch2} in a large bacterial population.
In this case, fluctuations of the  shared AI concentration $a$ due to the propagation of the MP noise average out over the population.
As a consequence, the AI concentration $a$ is related to the cell density $\rho$ by a one-to-one mapping and the fluctuations of MP abundances $\delta m^{(\rho)}_{i,t}$ are independent between cells.
With cell density $\rho$ treated as a fixed input value, the MP abundance $m^{(\rho)}_t$ becomes a stationary process, where the notation indicates fixed $\rho$.
The fraction of time that $m^{(\rho)}_t$ spends at any given MP abundance $m$ yields $p(m \vert \rho)$, the conditional probability of finding a concentration $m$ of MPs in a bacterium at cell density $\rho$.
As an encoding scheme, $p(m \vert \rho)$ specifies a feedback information channel $\rho \rightarrow a \leftrightarrows m \selfarrow$, which stochastically maps inputs $\rho$ onto outputs  $m$ via the shared intermediate $a$. \\
%
%
%
%
%
%
 \indent For bacteria such as \emph{V. harveyi}, the fluctuations of the quorum-sensing response are small with respect to the mean amplitude of the response~\cite{Campagna:2009uq,Teng:2010cr}.
This justifies our use of the small-noise approximation.
Accordingly, the encoding scheme $p(m \vert  \rho)$ will be well-described by a family of Gaussian distributions $\mathcal{N}\big(\overline{m}(\rho),\Sigma^2_m(\rho)\big)$, where $\overline{m}(\rho)$  and $\Sigma^2_m(\rho)$  are, respectively, the mean and the variance of $m^{(\rho)}$. 
This formulation of quorum sensing as a Gaussian information channel is shown schematically in Fig.~2, for a growing bacterial clone that engages in a series of plausible collective tasks, e.g. matrix synthesis, adhesion, antibiotic resistance, virulence and matrix degradation.
Despite this assumption of a stationary cell density, the encoding scheme $p(m \vert \rho)$  can still realistically describe quorum sensing in a growing colony with slow MP turnover, set only by dilution, with $\tau_m \approx  T_d \, \ln 2$ for cell-cycle period $T_d$ (see S1 Text).
Moreover, for abundant MP transcriptional factors (TFs), we can neglect the stochastic bindings of TFs to their cognate regulatory sequences as a source of noise in MP expression~\cite{Paulsson:2004uq,Tkacik:2012qf}.
This simplification permits us to consider a noise function $\sigma_m$ that only depends on the MP expression level at steady state $\overline{m}$ (see S1 Text).
In practice, noise in MP expression is more conveniently quantified via the steady-state MP Fano factor $F=\Sigma^2_m/\overline{m}$.
In the absence of feedbacks, the explicit expression of the Fano factor $F=\tau_m \sigma_m^2(\overline{m})/\overline{m}$ defines the Fano function $F^{(1)}(m)$, whose functional form can be inferred for simple models of gene expression~\cite{Taniguchi:2010kx,Munsky:2012dz}.
For convenience, we use this Fano function $F^{(1)}(m)$  instead of the noise function $\sigma_m$ to characterize noise in quorum sensing.

\begin{figure}[h]
\begin{center}
\includegraphics[width=0.85\textwidth]{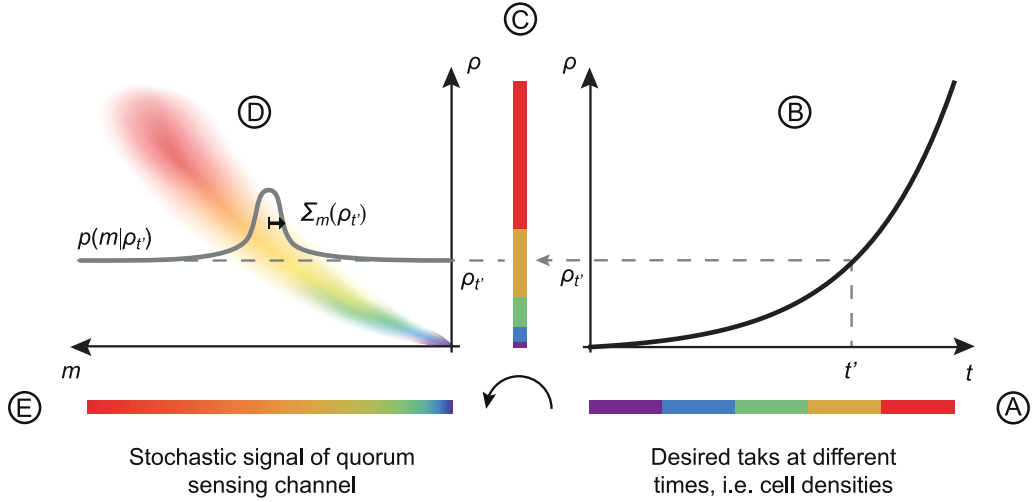}
\end{center}
\caption{
\textbf{Quorum-sensing information channel.}  
\textbf{A.} In this illustration, bacteria engage sequentially in five fictitious collective tasks, represented by five colors, that are homogeneously distributed in time. 
\textbf{B.} During the growth of a colony, the increasing cell density drives the quorum-sensing system.
\textbf{C.} To perform the desired tasks, bacteria need to resolve the five cell-density stages, whose probability is shaped by bacterial growth.
\textbf{D.} At fixed cell density, individual bacteria exhibit fluctuating levels of MPs, with mean $\overline{m}(\rho)$ and variance $\Sigma^2_m(\rho)$.
\textbf{E.} The smallest difference in cell density that a bacterium can resolve by reading out its fluctuating MP abundance specifies the resolution of the channel, defined as $\delta_\rho = \Sigma_m(\rho)/m'(\rho)$.
Thus, the information available to individual bacteria via quorum sensing depends both on the cell density dynamics and the channel resolution.
}
\end{figure}

\subsection*{Mutual information and information capacity}

%
%
In our dynamical model, a bacterium continuously monitors the AI concentration $a$, a proxy for the local cell density $\rho$, via its internal MP abundance $m$.
In principle, exploiting past measurements, i.e. consecutive uses of the quorum-sensing channel, could allow a bacterium to extract more information about $\rho$ than is available from instantaneous measurements.
However, in practice, a bacterium can only perform a simple temporal average of past measurements.
Specifically, in the case we consider, the instantaneous MP abundance, which controls the expression of downstream quorum-sensing genes, constitutes a long-time average of AI concentration measurements. 
Consequently, we quantify information transfer through the quorum-sensing channel --- cell density $\rightarrow$ AI concentration $\leftrightarrows$ MP abundance $\selfarrow$ (or $\rho \rightarrow a \leftrightarrows m \selfarrow$) --- by $I_{m,\rho}$, the MI between cell density $\rho$ and monitor abundance $m$. 
\\
%
%
%
%
%
\indent Given a specific instance of a quorum-sensing channel, i.e. for fixed functions $f_{\mathrm{ext}},f_m^{(1)},f_{\mathrm{int}},$ and Fano function $F^{(1)}(m)$, the optimal information transfer, called the information capacity $C$, is a characteristic of the channel~\cite{Shannon:1948fk,Cover:1991}.
In the small-noise regime, the capacity in bits is well-approximated by the integral
\begin{eqnarray}\label{eq:Capacity}
\tilde{C}=\log_2{ \int_{\rho_-}^{\rho_+} \frac{1}{\sqrt{2 \pi e} \delta_\rho} \, d\rho}\, ,
\end{eqnarray}
where $\delta_\rho = \Sigma_m(\rho)/\overline{m}\hspace{1pt}'(\rho)$ quantifies the smallest difference in cell density that a bacterium can resolve by reading out its MP abundance~\cite{Tkacik:2008ys}.
We refer to $\delta_\rho$ as the ``resolution'' of the quorum-sensing channel.
Importantly, the inverse of this quantity yields the input distribution for which the channel operates at capacity, $p(\rho) \propto 1/\delta_\rho$.
In other words, to fully exploit the capacity of the quorum-sensing channel, bacteria would need to grow such that the fraction of time they spend at a given cell density is inversely proportional to the resolution of the channel at that density.
To intuitively understand the above result, it helps to recognize that, at capacity, the output MP distribution satisfies 
 \begin{eqnarray*}\label{eq:InputCap}
q(m) \propto 1/\big(\delta_\rho \, \overline{m}\hspace{1pt}'(\rho)\big) = 1/\Sigma_m(m) \, ,
\end{eqnarray*}
where $\Sigma^2_m(m)$ is the MP variance for a mean abundance level $\overline{m}(\rho) = m$.
Hence, in the small-noise approximation, the capacity input distribution $p(\rho)$ is such that the MP output is distributed inversely with respect to its standard deviation $\Sigma_m$.
This is consistent with the intuition that efficient encodings of cell-density information should preferably utilize MP values associated with low output noise.\\
%
%
%
%
%
\indent For a fixed input distribution $p(\rho)$, one can contemplate optimizing $I_{m,\rho}$ by varying the encoding scheme  $p(m \vert \rho)$. 
For continuous variables, this is generally an ill-posed problem since $I_{m,\rho}$ diverges for deterministic mappings $p(m \vert  \rho) = \delta \left( m- \overline{m}(\rho) \right)$, where $\delta$ is the Dirac delta function.
This divergence is avoided by considering only quorum-sensing channels that always include a finite amount of noise.
Then, specifying Gaussian encoding schemes $\mathcal{N}\big(\overline{m}(\rho),\Sigma^2_m(\rho)\big)$ in terms of  $f_{\mathrm{ext}},f_m^{(1)},f_{\mathrm{int}},F^{(1)}$ allows us to write the MI $I_{m,\rho}$ as a functional of the external feedback $f_{\mathrm{ext}}$ and the internal feedback $f_\mathrm{int}$. 
In Methods, we present the small-noise approximation of $I_{m,\rho}$, denoted $\tilde{I}_{m,\rho}$, and formulate the optimization of $\tilde{I}_{m,\rho}$ over $f_{\mathrm{ext}}$ and $f_\mathrm{int}$ as a problem in the calculus of variations.
The solution of this variational problem for the quorum-sensing system yields the main results of our analysis.


\subsection*{Optimal mutual information in the small-noise approximation}

%
%
Optimizing the approximate MI $\tilde{I}_{m,\rho}$ over the external and internal feedbacks  $(f_{\mathrm{ext}},f_{\mathrm{int}})$ can be performed analytically for any quorum-sensing circuit satisfying the small-noise assumption (see S1 Text).
The optimal small-noise MI is
\begin{eqnarray}\label{eq:optMI}
\tilde{I}^\star_{m,\rho}= \frac{1}{2}\log_2 \left( \ln \left( \frac{m_+ f_{\mathrm{int},-}}{m_- f_{\mathrm{int},+}} \right) \int_{m_-}^{m_+} \frac{dm}{2 \pi e\, F^{(1)}\left( m \right) }  \right) \, ,
\end{eqnarray}
where $f_{\mathrm{int},-}$  and $f_{\mathrm{int},+}$  are the boundary values for the level of self-regulation via internal feedback.
Given a specified bare MP expression rate $f^{(1)}_m(c)$, the boundary values $f_{\mathrm{int},-}$ and $f_{\mathrm{int},+}$ can be deduced from the AI boundary values $(a_-,a_+)$ and MP boundary values $(m_-,m_+)$.
Alternatively, as the optimal MI \eqref{eq:optMI} does not  depend on $f^{(1)}_m(c)$ explicitly, we can consider $f_{\mathrm{int},-}$  and $f_{\mathrm{int},+}$ as boundary values on their own, indicating the level of self-regulation at the limits of low and high cell density.
If we constrain $f_{\mathrm{int},-}=f_{\mathrm{int},+}=1$, we find that the optimal internal feedback yields only a modest increase in information transmission over the capacity of the feedforward channel $a \rightarrow m$ (see S1 Text).
Without this constraint, the optimal information transfer $\tilde{I}^\star_{m,\rho}$ increases if self-regulation induces  MP synthesis in the low expression regime and represses MP synthesis in the high expression regime~\cite{Paulsson:2004uq}, i.e. if $f_{\mathrm{int},-}>1$ and $f_{\mathrm{int},+}<1$.
In \emph{V. harveyi}, the level of self-regulation of the monitor protein LuxR has been measured for both low and high  levels of expression~\cite{Teng:2011fk}, yielding $f_{\mathrm{int},-} \approx 2 $ and $f_{\mathrm{int},+} \approx 1/2$, which suggests a role for internal feedbacks in increasing information.\\
%
%
%
%
%
\indent Singularly, the optimal MI $\tilde{I}^\star_{m,\rho}$ is independent of the input distribution $p(\rho)$ and of the AI concentration range $(a_-,a_+)$.
This suggests that $\tilde{I}^\star_{m,\rho}$ is characteristic of the truncated detection channel  $a \rightarrow m \selfarrow$, taking the AI concentration $a$ as freely tuned input and without external feedback $f_{\mathrm{ext}}$, as opposed to the full quorum-sensing channel $\rho \rightarrow a \leftrightarrows m \selfarrow$.
In fact, $\tilde{I}^\star_{m,\rho}$ is equal to $\tilde{C}^\star$,  the optimal small-noise capacity of $a \rightarrow m \selfarrow$, obtained by varying the internal feedback $f_\mathrm{int}$ (see S1 Text).
To understand this equality, note that the deterministic dynamics of the AI concentration is shaped by the external feedback $f_{\mathrm{ext}}$.
By varying the external feedback $f_{\mathrm{ext}}$, one can deterministically match the distribution $q(c)$ to the resolution of $a \rightarrow m \selfarrow$ so that the quorum-sensing circuit operates at capacity $\tilde{C}[f_\mathrm{int}]$,  for any density distribution $p(\rho)$.
Then, one can always find the internal feedback $f_\mathrm{int}^\star$ for which $a \rightarrow m \selfarrow$ has the optimal capacity $\tilde{C}^\star$.
We stress that using a feedback  to tune an input to match a downstream channel is a general strategy to increase information transfer in biological systems.


\subsection*{Models for MP expression}

%
%
To further specify $\tilde{I}^\star_{m,\rho}$, we infer the Fano function $F^{(1)}(m)$ from models of genetic regulation.
A simple biologically relevant model posits that AIs freely diffuse across the bacterial membrane and induce the MP gene by regulating a TF.
In a slightly more complex example, the control of the MP gene by the TF is mediated by a small regulatory RNA (sRNA)~\cite{Lenz:2004mi},  which is also the target of the internal feedback.
In Fig.~3A and Fig.~3B, we  schematically represent a TF regulation model and an sRNA regulation model,  for which quorum-sensing information transmission can be computed~\cite{Mehta:2008uq,Mehta:2009ly}.
Both regulatory schemes can be modeled via Langevin equations, which prescribe the fluctuations of the MP at steady state (see S1 Text).
Solving the Langevin equations yields the simple Fano functions $F^{(1)}(m)=(1+b)/v$ for TF regulation and $F^{(1)}(m)=(1+b m/ m_\infty)/v$ for sRNA regulation, where the burst size $b$ is the average number of MPs  translated per mRNA transcript without sRNAs, $m_\infty = \max(m_+,m_+/f_{\mathrm{int},+})$ is the  MP concentration at saturation without self-inhibition, and $v$ is the cellular volume averaged over the bacterial population~\cite{Thattai:2001,Jost:2013kl}. 
These Fano functions are represented in Fig.~4E.\\
\indent The difference between the TF  Fano function and the sRNA Fano function can be understood intuitively.
Stochasticity in MP expression is mainly due to transcription noise, which is amplified by the burst size of the MP protein. 
For TF regulation, MP expression is downregulated by reducing the MP mRNA copy number.
As a result, the MP transcriptional noise arising from low mRNA copy number becomes substantial at low MP expression levels.
Compared with TF regulation, sRNA regulation downregulates MP expression by shortening the MP mRNA lifetime rather than reducing the MP mRNA copy number.
As the burst size $b$ equals the MP translation rate times the MP mRNA lifetime,  shorter MP mRNA lifetime yields an effective burst size that is smaller than the TF burst size $b$ (see S1 Text).
Hence, sRNA regulation reduces the Fano function at low MP expression levels.
This reduction of noise supposes that sRNA regulation operates in the regime where MPs are expressed above a base level ~\cite{Mehta:2008uq}, as is the case for LuxR in \emph{V. harveyi}~\cite{Teng:2010cr}.

\begin{figure}[h]
\begin{center}
\includegraphics[width=0.5\textwidth]{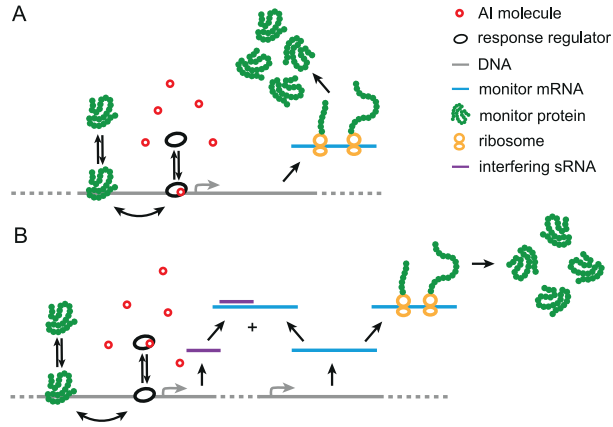}
\end{center}
\caption{
\textbf{Models for the regulation of the monitor protein (MP) expression.} 
\textbf{A.}  In the TF regulation model, AI molecules induce the production of MPs by allosterically regulating the transcription factor (TF), which only binds to its cognate DNA regulatory sequence when complexed with AI.
\textbf{B.} In the sRNA regulation model, a TF positively regulates a small regulatory RNA (sRNA) that  represses MP expression. 
In both models, the expression of MP is positively regulated by binding of AIs to the TF and MP proteins regulate their own expression.
For strong sRNA-mRNA pairing, sRNA regulation reduces the stochasticity in MP expression compared with TF regulation.
The internal feedback regulates the transcription of the MP mRNA for TF regulation, whereas it regulates the sRNA level for sRNA regulation.
}
\end{figure}


\subsection*{Optimal quorum-sensing response}

%
%
By adopting these simple models for genetic regulation, the knowledge of the bare AI output rate $f^{(1)}_m$ allows us to fully characterize the quorum-sensing response with optimal feedbacks (see S1 Text). 
As shown in the inset of Fig.~4A, $f^{(1)}_m$ is determined by a Hill activation curve with Hill coefficient $h=2$ and induction constant $K=15\mathrm{nM}$~\cite{Alon:2006oe}. 
For this choice of $f^{(1)}_m$, Fig.~4A and Fig.~4B shows the optimal time courses of AI concentration $\overline{a}\hspace{1pt}^\star(t)$ and of MP abundance $\overline{m}\hspace{1pt}^\star(t)$. 
These time courses show that the optimal response maximizes the fraction of time for which the quorum-sensing channel has high resolving power: 
The nonlinear time course $\overline{a}\hspace{1pt}^\star(t)$ reduces the effective range of AI concentrations to a small range around $K$, where MP expression is strongly inducible by AI detection via $f^{(1)}_m$. 
Independent of the choice of $f^{(1)}_m$ and in accordance with experimental observations~\cite{Teng:2010cr}, the quasi-linear increase $\overline{m}\hspace{1pt}^\star(t)$ exploits the full range of MP abundances to encode information about cell density.
In Fig.~4C and Fig.~4D, we exhibit the external feedback $f^\star_\mathrm{ext}$ and the internal feedback $f_\mathrm{int}^\star$ that achieve the optimal response.
The role of $f^\star_\mathrm{ext}$ is to transform the exponential growth function $\rho_t$ into the optimal time course of AI concentration $\overline{a}\hspace{1pt}^\star(t)$:
when increasing, $f^\star_\mathrm{ext}$ implements a  positive feedback to skip low and high AI concentration stages;
when decreasing, $f^\star_\mathrm{ext}$ implements a negative feedback that stabilizes the AI concentration around $K$.
The function $f_\mathrm{int}^\star$ similarly regulates the optimal MP output rate to yield the optimal time course of MP abundance $\overline{m}\hspace{1pt}^\star(t)$.
Interestingly, the nature of the optimal feedback for TF and sRNA regulation differ markedly at low MP abundance, where stochasticity in MP expression is substantially larger for TF regulation than for sRNA regulation.
For TF  regulation, $f_\mathrm{int}^\star$ is increasing and acts as a positive feedback to skip low MP abundance, whereas for sRNA regulation $f_\mathrm{int}^\star$ is decreasing to reduce stochasticity in this regime via negative feedback.

\begin{figure}[h]
\begin{center}
\includegraphics[width=0.5\textwidth]{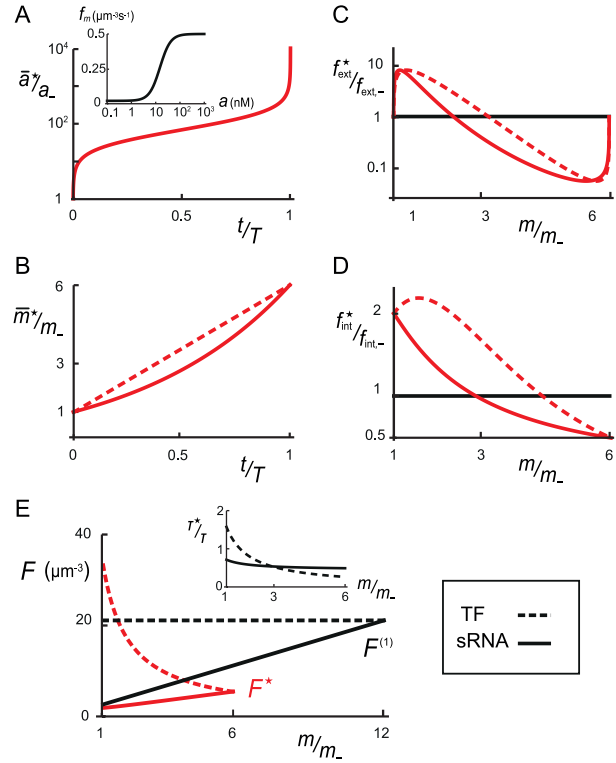}
\end{center}
\caption{
\textbf{Optimal quorum-sensing response.} 
TF regulation is shown by dashed curves and sRNA regulation by solid curves. 
\textbf{A.} The bare MP expression rate $f^{(1)}_m(c)$, in the absence of self-regulation of the MP, obeys a Hill function with $h=2$ (inset). Temporal dynamics of the optimal mean AI concentration and \textbf{B.} temporal dynamics of the optimal mean MP abundance. 
\textbf{C.} Optimal external feedback $f^\star_\mathrm{ext}$ with $f_- = a_-/(\tau_a \rho_-)$ and \textbf{D.} optimal internal feedback $f_\mathrm{int}^\star$.
\textbf{E.} Fano function  with optimal feedback $F^\star$ and without feedback  $F$ as functions of MP abundance $m$. 
The ratio of effective correlation times $\tau^\star/\tau = F^\star/F$ is shown inset.
In all panels, red curves indicate the optimal quorum-sensing response, black curves indicate the absence of all feedbacks $(f_{\mathrm{ext}}'=0,f_{\mathrm{int}}=1)$.
Parameter values: $\rho_+/\rho_-=10^4$, $a_-=0.1\mathrm{nM}$, $a_+=1\mathrm{mM}$, $m_-=100\mathrm{nM}$, $m_+=600\mathrm{nM}$, $f_{\mathrm{int},-}=2$, $f_{\mathrm{int},+}=1/2$, $K =15 \mathrm{nM}$, $h=2$, $b=20$ and $v= 1\mu\mathrm{m}^{3}$.
}
\end{figure}


\subsection*{Information increase via feedbacks}

%
%
To quantify the benefit of feedback to information transmission, we compare the MI for optimal feedbacks  $I^\star_{m,\rho}$ with the MI without feedback $I^{(1)}_{m,\rho}$.
As opposed to the analytical but approximate MIs  $\tilde{I}_{m,\rho}$ and $\tilde{I}^\star_{m,\rho}$, we compute numerically the MI $I^{(1)}_{m,\rho}$ for the feedforward channel $\rho \rightarrow a \rightarrow m$  and the MI $I_{m,\rho}^\star$ for the optimal feedback channel $\rho \rightarrow a \leftrightarrows m \selfarrow$.
Specifically, we use the analytical expressions for the optimal feedbacks to specify Gaussian information channels, but compute numerical MIs without the small-noise approximation (see S1 Text).
Table~1 gives the number of cell-density states that a bacterium can discriminate with or without feedback $(2^{I^\star_{m,\rho}}$ versus $2^{I^{(1)}_{m,\rho}}$ and with TF or sRNA regulation, for a fixed range of MP abundance $(m_-,m_+)$.
As the stochasticity in MP expression is less for sRNA regulation than for TF regulation, quorum-sensing channels with sRNA regulation can transmit more information than those with direct TF regulation~\cite{Jost:2013kl}.
For biologically-relevant values of the parameters, the total MI $I^\star_{m,\rho}$ barely exceeds $1$ bit of information for TF regulation ($\approx$ 2 states) but typically amounts to $2$ bits of information for sRNA regulation ($\approx$ 4 states).
For exponential growth, the external feedback only contributes marginally and most of the information increase is due to the internal feedback $f_\mathrm{int}^\star$ (see S1 Text).
To understand the benefit of $f_\mathrm{int}^\star$ to information transmission, we compare the Fano function, which is a convenient measure of noise, for optimal feedbacks $F^\star(m)$ to the Fano function without feedback $F^{(1)}(m)$:
\begin{eqnarray}\label{eq:Fstar}
\frac{F^\star(m)}{F^{(1)}(m)} = \left(1- m \frac{{f_\mathrm{int}^\star}'(m)}{f_\mathrm{int}^\star(m)} \right)^{-1}\, .
\end{eqnarray}
Thus, an increasing $f_\mathrm{int}^\star$ acts as a positive feedback and increases the Fano function, while a decreasing $f_\mathrm{int}^\star$ acts as a negative feedback and reduces the Fano function.
In Fig.~4E, we plot $F^\star$ and $F$ as functions of the MP abundance $m$. 
For TF regulation, $f_\mathrm{int}^\star$ is increasing at low MP abundance and incurs a loss of fidelity that approximatively cancels the information benefit achieved by more rapidly reaching  the enhanced resolution at higher MP abundance.
By contrast, for sRNA regulation, the steadily decreasing $f_\mathrm{int}^\star$ improves information transmission markedly (by $0.7$ bits), allowing a bacterium to resolve two additional cell-density states.
In both cases, the large regions of negative feedback ($f_\mathrm{int}^\star$ decreasing) shorten the effective correlation time of MP abundance, defined as $\tau^\star_m=\tau_m F^\star/F^{(1)}$.
Thus, the effective quorum-sensing timescale $\tau_m^\star$ becomes shorter than the division time $T_d/\ln2$, validating our use of the quasi-static approximation for dilution-limited MPs, with lifetime $\tau_m = T_d/\ln2$ (see S1 Text).\\

%
%
%
\begin{table}[!ht]
\begin{center}
\caption{
\textbf{Number of discernible cell-density stages.}}
\begin{tabular}{|c|c|c|}
\hline
Monitor regulation & No feedback & Optimal feedback\\
\hline
TF &$2$ ($1.2$ bits)& $2$ ($1.3$ bits)\\
\hline
sRNA & $2$-$3$ ($1.4$ bits) & $4$ ($2.1$ bits) \\
\hline
\end{tabular}
\begin{flushleft}
Information quantities are computed numerically using an analytical expression for the optimal feedback (see S1 Text).
Parameter values: $\rho_+/\rho_-=10^4$, $a_-=0.1\mathrm{nM}$, $a_+=1\mathrm{mM}$, $m_-=100\mathrm{nM}$, $m_+=600\mathrm{nM}$, $f_{\mathrm{int},-}=2$, $f_{\mathrm{int},+}=1/2$, $K =15 \mathrm{nM}$, $h=2$, $b=20$ and $v= 1\mu\mathrm{m}^{3}$
\end{flushleft}
\label{tab:label}
\end{center}
 \end{table}

%
%
%
\indent Focusing on our two simple models of regulation, what is the influence of the biological parameters on the quorum-sensing information transfer?
An obvious strategy to increase the quorum-sensing information transfer would be for bacteria to increase the output MP range $(m_-,m_+)$.
However, because MP abundances are plausibly bounded above and below to avoid toxicity to the cell or to maintain short response time, we treat $m_-$ and $m_+$ as boundary conditions.
Then, information transfer can be increased by widening the range $(f_{m,-},f_{m,+})$ of the bare output rate $f_m$, and by adjusting the resulting MP range $(m_-,m_+)$ via negative feedback, e.g. with self-repression at high MP level $f_{\mathrm{int},+}=m_+/ (\tau_m f_{m,+})<1$.
For a fixed MP range $(m_-,m_+)$, Fig.~5 illustrates the dependence of the MI $I_{\rho,m}^\star$ and of the MI increase $\Delta I_{\rho,m}^\star = I_{\rho,m}^\star-I^{(1)}_{\rho,m}$  on the burst size $b$ of MPs and on the level of  self-repression $f_{\mathrm{int},+}<1$ at AI saturation.
In particular, we plot the isoinformation curves, defined as the values $b$ and $f_{\mathrm{int},+}$ that yield the same optimal MI $I_{\rho,m}^\star$.
For TF regulation, i.e. $F^{(1)}=1+b$, the isoinformation curves are straight lines in the $(b,\ln f_{\mathrm{int},+})$-plane, as expected from expression \eqref{eq:optMI}.
Similarly, for sRNA regulation, i.e. $F^{(1)}(m)=1+b f_{\mathrm{int},+}m/m_+$, the isoinformation curves follow the predictions of \eqref{eq:optMI}, even with large burst sizes $(b\approx50)$ for which the small-noise approximation $\tilde{I}^\star_{\rho,m}$ underestimates $I^\star_{\rho,m}$.
Large burst sizes $b$ impact sRNA regulation much less than TF regulation since, in the sRNA case, the MP burst size  $b f_{\mathrm{int},+}m/m_+$ is smaller for small MP level $m$.
Moreover, the MI increase $\Delta I_{\rho,m}^\star$ in Fig.~5 reveals that internal feedbacks can improve information transmission for sRNA regulation but not for TF regulation.
Why doesn't internal feedback improve MI for TF regulation?
In principle, noise reduction in MP expression is possible for both TF regulation and sRNA regulation if bacteria can achieve strong enough feedbacks ( $f_{\mathrm{int},+} \ll 1$), and correspondingly large enough bare output rates $f_{m,+} = m_+/ (\tau_m f_{\mathrm{int},+})$.
However, there are biophysical limits to self-regulation efficacy and protein production rate, which limit the ability of feedback to control MP fluctuations, especially when gene expression noise is large.
For a realistic negative feedback and realistic burst size ($f_{\mathrm{int},-}=2$, $f_{\mathrm{int},+}=1/2$, and $b=20$), we find that sRNA regulation is in the regime where feedback can pay off in terms of increased information, but TF regulation is not due large gene expression noise at low MP abundance.
For sRNA regulation, the range of information transfer achievable (over $2$ bits) is significantly larger than the MI calculated in most circuits~\cite{Tkacik:2008ys,Cheong:2011fk,Tabbaa:2014}.
We believe that such an increase in achievable MI illustrates the power of assessing the benefit of feedback on MP time course  rather than on static MP abundance.

\begin{figure}[h]
\begin{center}
\includegraphics[width=0.5\textwidth]{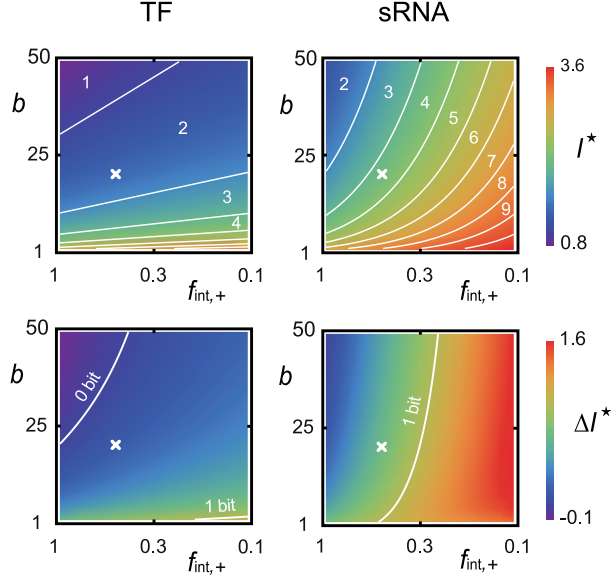}
\end{center}
\caption{
\textbf{Optimal MI and optimal MI increase.} 
Dependence of the optimal MI $I^\star$ and the optimal MI increase $\Delta I^\star$ (in bits)  on the burst size $b$ and on the level of self-repression $f_{\mathrm{int},+}$ (in logarithmic scale) for both TF and sRNA regulations. 
The Xs indicate the values $f_{\mathrm{int},+}=1/2$ and $b=20$ for which Table~1 was computed.
In the top panels, the white curves are isoinformation curves separating regions where the optimal quorum-sensing channel can discriminate the indicated number of cell-density ranges.
In the bottom panels, the white curves represent parameters for which feedbacks cannot improve information transfer ($0$ bits) or can double the number of distinguishable cell density ranges ($1$ bit).
Note that, in both cases, a decrease in $f_{\mathrm{int},+}$ has to be matched by a larger bare MP output rate to ensure the boundary condition $\tau_m f_{\mathrm{int},+} f^{(1)}_m(a_+) = m_+$.
Parameter values: $\rho_+/\rho_-=10^4$, $a_-=0.1\mathrm{nM}$, $a_+=1\mathrm{mM}$, $m_-=100\mathrm{nM}$, $m_+=600\mathrm{nM}$, $f_{\mathrm{int},-}=2$, $K =15 \mathrm{nM}$, $h=2$ and $v = 1\mu\mathrm{m}^{3}$.
}
\end{figure}


\section*{Discussion}


As an information channel, the quorum-sensing system encodes input cell density $\rho$ into output MP abundance $m$ via the intermediary of AI concentration $a$.
While the external AI  concentration $a$ determines the production rate of self-regulating MPs, the per-cell AI-output rate depends on the concentration $m$ of the internal MP, thus establishing a channel of the form $\rho \rightarrow a \leftrightarrows m \selfarrow$.
To assess the information benefit of feedbacks from MP level  to AI production and to MP expression, we optimized the small-noise MI $\tilde{I}_{m,\rho}$ over both the external and internal feedbacks.
In our model, external and internal feedbacks actually decouple to increase information transmission: the external feedback adjusts the time course of the AI concentration to the inherent noise of the detection channel $a \rightarrow m \selfarrow$~\cite{Tkacik:2009cr,Walczak:2010dq,Tkacik:2012qf}, while the internal feedback optimizes the information capacity of the detection channel~\cite{Yu:2008vn}.
This result constitutes a generalized form of ``histogram equalization'' for a noisy information channel.
For a channel with uniform output noise, i.e. constant $\Sigma_m$, and fixed output bandwidth $(m_-,m_+)$, specifying the input/output mapping $\overline{m}\hspace{1pt}^\star(\rho)$ as the scaled input cumulative distribution function 
\begin{eqnarray}\label{eq:HistEqual}
\overline{m}\hspace{1pt}^\star(\rho) =m_- + (m_+ - m_-) P(\rho) \, , \quad P(\rho) = \int^{\rho}_{\rho_-} p(u) \, du \, ,
\end{eqnarray}
optimizes information transfer in the small-noise approximation.
Indeed, for uniform output noise, optimizing the MI $I_{\rho,m}$ amounts to maximizing  the output entropy of $p(m)$.
In this regard, the mapping $\overline{m}\hspace{1pt}^\star(\rho)$ defined by \eqref{eq:HistEqual} transforms the input distribution $p(\rho)$ into the uniform output distribution $p(m)$ over $(m_-,m_+)$, thereby maximizing the output entropy: This is classical histogram equalization that ensures a uniform use of the output bandwidth~\cite{Laughlin:1981zr}.
For a noisy channel, direct histogram equalization fails to optimize information transfer in general.
Rather than using the output bandwidth uniformly, one has to preferentially exploit the output bandwidth where the output noise is low.
Our information analysis indicates the optimal way to allocate bandwidth in the small-noise approximation.
As in histogram equalization, the information transfer is optimized by adjusting the shape of smooth curves, namely the time-courses of AI concentration and MP abundance.
However, unlike histogram equalization, these feedback-mediated adjustments not only depend on the input statistics but also on the noise characteristics of the encoding channel.
As a result, determining the optimal adjustments is a problem in the calculus of the variations and the optimal information transfer will depend on the noise properties of the channel.
Specifically, tuning the quorum-sensing feedbacks yields contrasting benefits for two biologically relevant models of MP genetic regulation:
for sRNA-based regulation, optimal feedback can double the number of distinguishable cell-density ranges, while feedbacks are only marginally beneficial for TF-based regulation.
In both cases, for exponential growth, the quorum-sensing circuit operates close to capacity at constant AI output rate (i.e. no external feedback).
Thus, the only feedback we find that substantially increases MI is internal feedback on MP levels for sRNA-based regulation.\\
%
%
%
%
%
%
\indent  During the growth of a colony, quorum-sensing bacteria activate different programs of gene expression based on their MP levels.
If only cells that respond appropriately to cell density survive, the gain of fitness is theoretically equal to the MI between MP abundance and cell density~\cite{Rivoire:2011ly,Kussell:2005}.
This formal identification of MI with fitness gain can account for the optimization of MI as the result of competition among bacterial strains.
In this context,  a natural strategy for a bacterium to increase its MI is to use a genetic circuit that reduces noise by averaging many consecutive measurements.
In our model, because the MP lifetime is only dilution limited, the MP abundance performs a long-time average of the discrete random events inherent to signal transduction and gene expression.
Such a temporal average allows a bacterium to exploit temporal correlations in the input to maximize available information. 
In fact, temporal averaging is the only possible memory management in quorum sensing since a bacterium cannot store past molecular abundances as distinct time-stamped values.
Moreover, the averaging time should be as long as possible for optimal noise filtering, but short enough to accurately track the changes in cell density relayed by the AI signal.  
By tuning internal feedbacks, a bacterium can adjust the dynamics of its quorum-sensing response to achieve the optimal trade-off between noise reduction and input tracking ability.
Biologically, such internal feedbacks suppress molecular fluctuations dynamically, which generally requires fast feedbacks involving many molecular events, i.e. high turnover rates for intermediary molecules~\cite{Paulsson:2004uq}.
In our case, effective feedback can be achieved by expressing MP mRNA at a maximal rate, while increasing the expression rate of the complementary sRNA, effectively increasing mRNA/sRNA turn-over rates.
In this regard, the self-regulation of a slow monitor protein via fast sRNA regulation appears as a trademark of high-MI signal tracking by a genetic circuit~\cite{Jost:2013kl}.\\
%
%
%
%
%
\indent In reality, many bacterial species use multiple AIs, multiple MPs, as well as multiple mechanisms of gene regulation~\cite{Ng:2009nx,Miller:2001oq,Nadal-Jimenez:2012hc}.
In addition, bacteria grow in complex communities, such as biofilms, that comprise many species that possibly communicate and compete via quorum sensing~\cite{Fuqua:1996bh}.
Our information-theoretic approach can be extended to address these real-world considerations. 
If the communal AI-concentration signals self-average in interacting bacterial populations, external feedbacks can always increase information transfer by tuning the AI concentrations to the specifics of the detection channels. 
For a densely packed biofilm, the constant reshaping of the AI diffusion volume due to bacterial growth can be modeled by variable AI diffusion times.
Corrugated geometries or complex diffusive environments can lead to inhomogeneous AI concentrations~\cite{Youk:2014fk}. 
In such cases, the local AI concentration may be plagued by slow fluctuations due to the stochastic AI output from a small number of neighboring cells, each of which is subject to slow internal monitor protein fluctuations and therefore slow AI output fluctuations.
These AI fluctuations represent a form of extrinsic noise.
Such irreducible noise in the AI signal restrains the ability of feedback to reshape the AI distribution $p(c)$ by tuning the AI concentration.
In particular, it generally becomes impossible for $p(c)$ to match the capacity input distribution of the detection channel $c \rightarrow m$. 
Therefore, the optimal MI will no longer achieve channel capacity.
As for bacteria where MP expression is controlled both by sRNA regulation and TF regulation, such as \emph{V. harveyi}~\cite{Rutherford:2012fk}, our analysis suggests there is no benefit to using both modes of regulation simultaneously.
Rather, TF regulation should be active at the earliest stage of quorum sensing, where the noise level is high, to encode one bit of information.
Triggered by this one bit, sRNA regulation can then take over for faithful information processing at higher cell densities.\\
%
%
%
%
%
\indent More generally, the feedback structure of the quorum-sensing system $x \rightarrow y \leftrightarrows z \selfarrow$ is ubiquitous in multi-agent systems that need to monitor their resources in order to synchronize their activity, be it a multicellular community, a developing organ~\cite{Hietakangas:2009ff}, or a computing network system~\cite{Murray:2007}.  
As a strategy, agents in such systems infer a changing resource density from a common self-generated signal, whose dynamics is driven by the process to be monitored.
In that respect, our analysis is relevant to a wide range of statistical systems that track an  extrinsic driver implicitly, via an endogenously generated signal.


\section*{Methods}


We first formulate the encoding scheme $p(m \vert \rho)\sim \mathcal{N}(\overline{m}(\rho),\Sigma^2_m(\rho))$, expressing $\overline{m}(\rho)$ and $\Sigma^2_m(\rho)$ as functionals of $f_{\mathrm{ext}},f_m^{(1)},f_{\mathrm{int}},F^{(1)}$.
This requires the analysis of the coupled dynamics of the stationary process $(c^{(\rho)},  \lbrace m_i^{(\rho)}   \rbrace )$, which describes quorum sensing in a colony of bacteria held at fixed cell density $\rho$. \\
%
%
%
%
%
%
\indent
At any time in such a colony, the AI concentration and the MP abundances fluctuate around their respective mean values $\overline{a}(\rho)$ and $\overline{m}(\rho)$.
This defines $(\overline{a}(\rho),\overline{m}(\rho))$ as the stable fixed point of the deterministic versions of equations \eqref{eq:qsStoch1} and \eqref{eq:qsStoch2} (i.e. with $\sigma_m  = 0$).
The fixed-point condition imposes the self-consistent relations
\begin{equation}\label{eq:meanField}
 \overline{a} = \tau_a \rho f_{\mathrm{ext}} (\overline{m}) \quad \mathrm{and} \quad    \overline{m}=\tau_m f_m^{(1)}( \overline{a}) f_{\mathrm{int}}( \overline{m}) \, , 
\end{equation}
which implicitly define the cell density-AI concentration mapping $\overline{a}(\rho)$ and the cell density-MP abundance mapping $\overline{m}(\rho)$.
We impose the constraints that $\overline{a}(\rho)$ and $\overline{m}(\rho)$ are differentiable increasing mappings, thus avoiding multistability~\cite{Angeli:2004}, which is known to impair information transmission~\cite{Tkacik:2012qf}.
In particular, these constraints allow $f_{\mathrm{ext}}$ and $f_\mathrm{int}$ to be non-monotonic, while imposing that $f_\mathrm{int}'(m)/f_\mathrm{int}(m)<1/m$ (see equation \eqref{eq:Fstar}.
It is convenient to consider the MI as a function of the cell density-AI concentration mapping $\overline{a}(\rho)$ and the AI concentration-MP abundance mapping $\overline{m}(\rho)$ instead of as a function of $f_{\mathrm{ext}}$ and $f_\mathrm{int}$.
Thus, the optimization of the MI is carried out over the space of increasing functions $ \overline{a}(\rho)$ and $ \overline{m}(\rho)$ satisfying $ \overline{a}(\rho_-)=a_-$ and $\overline{a}(\rho_+)=a_+$, as well as $ \overline{m}(\rho_-)=m_-$ and $ \overline{m}(\rho_+)=m_+$.\\
%
%
%
%
%
%
\indent In the small-noise approximation, the fluctuations $( \delta c^{(\rho)} , \lbrace \delta m_i^{(\rho)} \rbrace )$ around $(\overline{a}(\rho),  \overline{m}(\rho))$ satisfy the linearized versions of equations \eqref{eq:qsStoch1} and \eqref{eq:qsStoch2}. 
The integral expression for the covariance matrix of the stationary process $\lbrace \delta m_i^{(\rho)} \rbrace$ yields the variance $\Sigma^2_m(\rho) = \langle \delta m_i^{(\rho)} \, \delta m_i^{(\rho)} \rangle$ of the MP abundance in a bacterium at cell density $\rho$.
When the shared AI concentration is self-averaging, i.e. $\Sigma^2_a(\rho) = \langle \delta a{(\rho)} \, \delta a{(\rho)} \rangle=0$, it is possible to obtain a simple expression for $\Sigma^2_m(\rho)$, revealing that the noise amplitudes are modified by the feedback mechanisms through the first derivatives $f_{\mathrm{ext}}'$ and $f_\mathrm{int}'$.
This implies that, as a functional of $\overline{a}(\rho)$ and $\overline{m}(\rho)$, $\Sigma^2_m(\rho)$ depends not only on $\overline{a}(\rho)$ and $\overline{m}(\rho)$, but also on the sensitivity of the AI concentration  $\overline{a}\hspace{1pt}'(\rho)$ and of the MP abundance $\overline{m}\hspace{1pt}'(\rho)$ with respect to the cell density $\rho$ (see S1 Text).\\
%
%
%
%
%
%
\indent We employ a variational method to optimize $\tilde{I}_{m,\rho}$ over the feedback functions $f_{\mathrm{ext}}$ and $f_\mathrm{int}$, while holding $f_m^{(1)}$ and $F^{(1)}$ fixed.
In the small-noise approximation,  the trajectories of  $m_t$ fluctuate closely around their deterministic mean $\overline{m}(t)$ during colony growth.
Assuming a fixed time course for the growth of cell density, we may neglect the contribution of the small transient fluctuations $\delta m_t$ in shaping the AI concentration distribution $q(c)$ and the MP abundance distribution $q(m)$.
The cell density-MP abundance mapping $\overline{m}(\rho)$ deterministically maps the input distribution $p(\rho)$ onto the output distributions $q(m) = p(\rho) /  \overline{m}\hspace{1pt}'(\rho)$. 
Thus, the small-noise MI $\tilde{I}_{m,\rho}$ can be written
\begin{eqnarray}\label{eq:snMI}
\tilde{I}_{m,\rho}
=
 H_\rho + \int_{\rho_-}^{\rho_+} p(\rho) \log_2{\left( \frac{1}{\sqrt{2\pi e}} \frac{ \overline{m}\hspace{1pt}'(\rho)}{\Sigma_m(\rho)}\right)} \, d\rho \, ,
\end{eqnarray}
where  $H_\rho$ is the continuous entropy associated with $p(\rho)$ (see S1 Text).
As the approximation $q(m) = p(\rho) /  \overline{m}\hspace{1pt}'(\rho)$ neglects noise, $\tilde{I}_{m,\rho}$ underestimates $H_m$ the entropy of $q(m)$, yielding only a lower bound to the true MI $\tilde{I}_{m,\rho}$.
However, expression \eqref{eq:snMI} has a clear interpretation.
The larger the sensitivity-to-noise ratio $\overline{m}\hspace{1pt}'(\rho)/\Sigma_m(\rho)$, the more faithful the encoding becomes at local density $\rho$:
the ratio $\delta_\rho = \Sigma_m(\rho)/\overline{m}\hspace{1pt}'(\rho)$, referred as the resolution of the quorum-sensing channel, quantifies the smallest difference in cell density that a bacterium can resolve by reading out its MP abundance.
The logarithmic contribution of this resolution to the overall information is weighted by the probability $p(\rho)$, which captures the fraction of time the colony spends at density $\rho$. \\
%
%
%
%
%
%
\indent Since $\overline{m}\hspace{1pt}'(\rho)/\Sigma_m(\rho)$ implicitly depends on the mean mappings $\overline{a}(\rho)$, $\overline{m}(\rho)$, and their first derivatives, the optimization of $\tilde{I}_{m,\rho}$ over $\overline{a}$ and $\overline{m}$, that is over $f_{\mathrm{ext}}$ and $f_\mathrm{int}$, becomes a problem of the calculus of variations.
If there are optimal mean mappings $\overline{a}\hspace{1pt}^\star(\rho)$ and $\overline{m}\hspace{1pt}^\star(\rho)$, these necessarily define a stationary path solving the Euler-Lagrange equations of the variational problem.
Moreover, a simple analysis of $\tilde{I}_{m,\rho}$ confirms that the stationary path $\overline{a}\hspace{1pt}^\star$ and $\overline{m}\hspace{1pt}^\star$ actually gives a local maximum of the MI (see S1 Text).
The Euler-Lagrange equations corresponding to our quorum-sensing model are nonlinear second-order equations that depend on the functional parameters $f_m^{(1)},F^{(1)}$.
As such, their analytical resolution is in principle a formidable task. 
However, the variational optimization of $\tilde{I}_{m,\rho}$ can be carried out analytically in our case.


\section*{Acknowledgments}

We thank William Bialek, Bonnie Bassler and Curt Callan for many insightful discussions.

\end{document}